\documentclass[a4paper,11pt]{article}
\pdfoutput=1
\usepackage{amssymb}
\usepackage{epsfig}
\usepackage{graphicx}

\makeatletter

\@addtoreset{equation}{section}
\makeatother
\def\be{\begin{equation}}
\def\ee{\end{equation}}
\def\bea{\begin{eqnarray}}
\def\eea{\end{eqnarray}}

\def\({\left(}
\def\){\right)}
\def\<{\left<}
\def\>{\right>}

\def\tr{{\mbox{tr}}}
\def\be{\begin{equation}}
\def\ee{\end{equation}}
\def\bea{\begin{eqnarray*}}
\def\eea{\end{eqnarray*}}
\def\ben{\begin{eqnarray}}
\def\een{\end{eqnarray}}
\def\({\left(}
\def\){\right)}
\def\<{\left<}
\def\>{\right>}
\def\!{\right|}
\def\|{\left|}

\def\[{\left[}
\def\]{\right]}

\def\+{\bar}
\def\mb{\mathbb}
\def\tr{{\mbox{tr}}}

\def\La{{\cal{L}}}

\begin{document}

\setlength{\unitlength}{1mm}

\pagestyle{empty}
\vskip-10pt
\vskip-10pt
\hfill 
\begin{center}
\vskip 3truecm
{\Large \bf
Partially twisted superconformal M5 brane \\
in R-symmetry gauge field backgrounds}
\vskip 2truecm
{\large \bf
Dongsu Bak$^a$ and  Andreas Gustavsson$^{a,b}$}
\vspace{1cm} 
\begin{center} 
\it  a) Physics Department, University of Seoul,  Seoul 130-743, Korea\\
b) School of Physics, Korea Institute for Advanced Study, Seoul 130-012, Korea
\end{center}
\vskip 0.7truecm
\begin{center}
(\tt dsbak@uos.ac.kr, agbrev@gmail.com)
\end{center}
\end{center}
\vskip 2truecm
{\abstract{We obtain the action for a curved superconformal abelian M5 brane with the background R-symmetry gauge field turned on. We then restrict ourselves to superconformal M5 brane on a sphere times flat Minkowski space. We choose R-symmetry $SO(1,4)$ instead of $SO(5)$, which enables us to partially twist on Minkowski space and replace it by some curved Lorentzian manifold. We obtain M5 brane actions on $M_{1,1} \times S^4$ and $M_{1,2} \times S^3$ where actions and all fields, including the background gauge field, are real. Dimensional reduction along time gives real 5d SYM actions with nonabelian generalizations.}}

\vfill
\vskip4pt
\eject
\pagestyle{plain}

\section{Introduction}
A superconformal M5 brane can be put in a generic conformal supergravity background \cite{Bergshoeff:1999db}. The corresponding supergravity background fields in the dimensionally reduced 5d SYM theory has been analyzed in \cite{Cordova:2013bea} following the approach of \cite{Festuccia:2011ws}. Using this result, 5d SYM theories on $\mb{R}^{3} \times S^2$ \cite{Cordova:2013cea} and on $\mb{R} \times S^4$ \cite{Kim:2014kta} have been obtained. However the corresponding Lagrangian of the abelian M5 brane has not been obtained\footnote{A related question was addressed in \cite{Gran:2014lia}. Here the abelian M5 brane Lagrangian was obtained on geometries of the form $\mb{R}^{1,1} \times M_4$ where a partial topological twist of Donaldson-Witten type was performed on $M_4$.}, perhaps due to the belief that no such Lagrangian can be written down because of the selfdual tensor field in 6d. However, by also including the wrong chirality tensor field as a decoupled spectator field, we can write down a superconformal Lagrangian in 6d. But another reason that no 6d Lagrangian has been obtained in the literature might be the following. In the applications to the AGT correspondence \cite{Alday:2009aq} and the 3d-3d correspondence \cite{Cordova:2013cea}\footnote{The many original papers that proposed the 3d-3d correspondence can be found in the reference list of \cite{Cordova:2013cea}.}, we like to put M5 brane on $\mb{R}^p \times S^{6-p}$ for $p=2$ and $p=3$ respectively, and then perform a partial topological twist with the $SO(5)$ R symmetry that enables us to put the theory on $M_p \times S^{6-p}$ for a general $p$-manifold $M_p$. The theory being topological on $M_p$ means that we can scale the size of $M_p$ without affecting any observables in the theory. By taking the size to be small we obtain a dimensionally reduced SYM theory on $S^{6-p}$. By taking the size to be large we obtain a theory on $M_p$. These theories will be equivalent thanks to the topological property of the theory on $M_p$. 

However, one obstacle in carrying out such a computation explicitly is that no M5 brane Lagrangian can exist in Euclidean signature with real fermions. If we consider the theory in Lorentzian signature, we should, for $p=2$ consider the manifold $\mb{R}^{1,1} \times S^4$. However, as was mentioned in \cite{Anderson:2013hpa}, we cannot twist this theory partially on $\mb{R}^{1,1}$ if the R symmetry group is $SO(5)$. 

In this paper we propose to solve this problem by instead taking the R symmetry group to be $SO(1,4)$ \cite{Hull:1999mt,Hull:2014cxa}.\footnote{We use a signature convention such that $SO(1,4)$ refers to the group of transformations that leaves the metric diag$(-1,+1,+1,+1,+1)$ invariant. We then also refer to this space as $\mb{R}^{1,4}$ or as a space of signature $(1,4)$.} This enables us to twist an $SO(1,p-1)$ subgroup with the Lorentz group $SO(1,p-1)$ on $\mb{R}^{1,p-1}$. We may then put the theory on a general Lorentzian $p$-manifold $M_{1,p-1}$ times $S^{6-p}$. 

For $p=1,2,3$ we can find solutions for the background gauge potential, and the full M5 brane Lagrangian becomes real in Lorentzian signature. It is required that the bosonic part of the Lagrangian is real in order to have a unitarity of the theory \cite{Witten:1989ip}. What is problematic though, is that with $SO(1,4)$ R symmetry we have an indefinite kinetic energy for the scalar fields. But this kind of problem might be cured by finding a suitable integration cycle where the path integral is convergent. For more details we refer to section $3$.

We will also perform dimensional reduction along time. This will perhaps justify our choice of R symmetry group as $SO(1,4)$ a bit further. After that we dimensionally reduce flat M5 brane with $SO(1,4)$ R symmetry along time, we find precisely the 5d SYM that has global symmetry $SO(5) \times SO(1,4)$ that also can be obtained by dimensionally reducing 10d SYM with $SO(1,9)$ global symmetry by reduction along time and 4 spatial directions. The latter approach has been used in for example \cite{Pestun:2007rz} to derive a SYM Lagrangian on a four-sphere from 10d SYM with real fermions. 

In this paper we will restrict ourselves to just turning on the supergravity background gauge field that is associated with the $SO(1,4)$ R symmetry. Thus we will put all the other supergravity fields to zero. Our restriction has the unfortunate limitation that we cannot consider squashed spheres as these require other background fields also being turned on. The AGT-like correspondences of course become much more interesting if one can include an additional squashing parameter in the correspondence. We plan to return to this problems in a future publication.

\section{Abelian 6d theory with $SO(1,4)$ R symmetry group}
In the introduction we have motivated why we like to study 6d $(2,0)$ theory with $SO(1,4)$ R symmetry group. This can be thought of as embedding a Lorentzian M5 brane into 11 dimensional space with signature $(2,9)$. Let us now work out the supersymmetry transformations assuming $SO(1,4)$ R-symmetry group. We start by considering M5 brane on flat $\mb{R}^{1,5}$. We use 11d gamma matrices that we split as $\Gamma^M$ ($M=0,...,5$) and $\hat\Gamma^A$ ($A = 0',...,4'$) and define the 6d chirality matrix $\Gamma = \Gamma^{012345}$. The spinor and the supersymmetry parameter have opposite 6d chiralities. We choose the convention
\bea
\Gamma \psi &=& \psi\cr
\Gamma \epsilon &=& -\epsilon
\eea
The 11d Majorana conditions (or, equivalently, the 6d $SO(1,4)$-Majorana conditions) for these chiral spinors read
\bea
\bar\psi &=& \psi^T C\cr
\bar\epsilon &=& \epsilon^T C
\eea
where $\bar\psi = \psi^{\dag}  \Gamma^0 \hat\Gamma^{0'}$. We find that the following supersymmetry variations
\bea
\delta \phi^A &=& \bar\epsilon \hat\Gamma^A \psi\cr
\delta B_{MN} &=& i \bar\epsilon \Gamma_{MN} \psi\cr
\delta \psi &=& -\frac{i}{12}\Gamma^{MNP} \epsilon H_{MNP} + \Gamma^M \hat\Gamma_A \epsilon \partial_M \phi^A
\eea
close on-shell, 
\bea
[\delta_{\eta},\delta_{\epsilon}] \phi^A &=& -2 \bar\epsilon \Gamma^P \eta \partial_P \phi^A\cr
[\delta_{\eta},\delta_{\epsilon}] B_{MN} &=& -2 \bar\epsilon \Gamma^P \eta H_{PMN}\cr
[\delta_{\eta},\delta_{\epsilon}] \psi &=& -2 \bar\epsilon\Gamma^P\eta \partial_P \psi + \frac{3}{4} \bar\epsilon\Gamma^P\eta \Gamma_P \Gamma^M \partial_M \psi\cr
&& - \frac{1}{4} \bar\eta\Gamma^M \hat\Gamma^A\epsilon\Gamma_M\hat\Gamma_A\Gamma^N\partial_N\psi
\eea
To obtain the closure relation for the fermion we have used the Fierz identity that we have collected in Appendix \ref{D}. For closure we must use the fermionic equation of motion
\bea
\Gamma^M \partial_M \psi &=& 0
\eea
Let us notice that 
\bea
(\bar\epsilon \Gamma^M \eta)^* &=& -(-1)^{\frac{q(q+1)}{2}} \bar\epsilon\Gamma^M\eta
\eea
where $q$ counts the number of timelike components in the R symmetry group $SO(q,5-q)$.
In particular then, while we have that $\bar\epsilon\Gamma^M\eta$ is purely imaginary for $SO(5)$ R symmetry, we find that $\bar\epsilon\Gamma^M\eta$ becomes real for $SO(1,4)$ R symmetry. This explains why we do not get the usual factor of 
$i$ in the closure relations, such as $\sim 2i \bar\epsilon\Gamma^M \eta \, \partial_M \phi^A$ as we get when the R symmetry is $SO(5)$. 

By using the 11d Majorana condition, one can see that $\delta \phi^A$ and $\delta B_{MN}$ are real, and that the variation $\delta \psi$ again satisfies the 11d Majorana condition. We notice that the factors of $i$ sit at different places compared to the more commonly used supersymmetry transformations for the $(2,0)$ theory that has $SO(5)$ R symmetry group. 
 
As usual, from $\Gamma\epsilon = -\epsilon$, we can find that the gauge field part of the above supersymmetry variations can be also written in the form\footnote{The dots represent the scalar field part.} 
\bea
\delta H_{MNP}^+ &=& \frac{i}{2} \bar\epsilon \Gamma^Q \Gamma_{MNP} \partial_Q \psi\cr
\delta H_{MNP}^- &=& 0\cr
\delta \psi &=&  -\frac{i}{12}\Gamma^{MNP} \epsilon H^+_{MNP} + ...
\eea
where we define
\bea
H_{MNP}^\pm &=& \frac{1}{2} \(H_{MNP} \pm \frac{1}{6} \epsilon_{MNP}{}^{UVW} H_{UVW}\)
\eea
This means that $H^-_{MNP}$ is not part of the tensor multiplet, but we include it in order  to write down a neat supersymmetric Lagrangian, which is given by
\bea
\La &=& -\frac{1}{24} H^{MNP} H_{MNP} + \frac{1}{2} \partial^M \phi^A \partial_M \phi_A - \frac{1}{2} \bar\psi \Gamma^M \partial_M \psi
\eea
First we notice that the whole Lagrangian is real. In particular we have
\bea
(\bar\psi \Gamma^M \partial_M \psi)^{\dag} &=& \bar\psi \Gamma^M \partial_M \psi
\eea
up to a boundary term produced by an integration by parts. Second, we notice that the gauge potential kinetic term and the scalar field kinetic term cannot both have the right sign simultaneously. However, for the kinetic term of the scalar fields, we also need to remember that the signature of the R symmetry group is $SO(1,4)$ which means that it is never possible for all the five scalar fields to have the right sign of the kinetic term. It is therefore the most natural to assign the gauge potential the right sign kinetic term, and then $\phi^{a'}$ for $a' = 1',2',3',4'$ will have the wrong sign kinetic term. 

\subsection{$SO(4,1)$ R symmetry group}
For completeness, we also work out the supersymmetry variations with $SO(4,1)$ R symmetry group, which corresponds to $(5,6)$ signature in $11$ dimensions.
We have the supersymmetry variations
\bea
\delta \phi^A &=& i\bar\epsilon \hat\Gamma^A \psi\cr
\delta B_{MN} &=& i \bar\epsilon \Gamma_{MN} \psi\cr
\delta \psi &=& \frac{1}{12}\Gamma^{MNP} \epsilon H_{MNP} + \Gamma^M \hat\Gamma_A \epsilon \partial_M \phi^A
\eea
where the Dirac conjugation $\bar\chi$ is now defined by $\chi^\dagger \Gamma^0 \hat\Gamma^{1'2'3'4'}$ in this $SO(4,1)$ theory.
By using the 11d Majorana condition in this signature, one can see that $\delta \phi^A$ and $\delta B_{MN}$ are real, and that the variation $\delta \psi$ again satisfies the 11d Majorana condition. Closure relations are
\bea
[\delta_{\eta},\delta_{\epsilon}] \phi^A &=& -2i \bar\epsilon \Gamma^P \eta \partial_P \phi^A\cr
[\delta_{\eta},\delta_{\epsilon}] B_{MN} &=& -2i \bar\epsilon \Gamma^P \eta H_{PMN}\cr
[\delta_{\eta},\delta_{\epsilon}] \psi &=& -2i \bar\epsilon\Gamma^P\eta \partial_P \psi + 
\frac{3i}{4} \bar\epsilon\Gamma^P\eta \Gamma_P \Gamma^M \partial_M \psi\cr
&& - \frac{i}{4} \bar\eta\Gamma^M\hat\Gamma^A\epsilon\Gamma_M \hat\Gamma_A\Gamma^N\partial_N\psi
\eea
That is, we have on-shell closure on the fermionic equation of motion
\bea
\Gamma^M \partial_M \psi &=& 0
\eea
 
The supersymmetric Lagrangian is
\bea
\La &=& -\frac{1}{24} H^{MNP} H_{MNP} - \frac{1}{2} \partial^M \phi^A \partial_M \phi_A + \frac{i}{2} \bar\psi \Gamma^M \partial_M \psi
\eea

This theory can be obtained from the above theory in signature $(2,9)$ by the following map, 
\bea
&& \Gamma^M \rightarrow \Gamma^M \cr
&& \hat\Gamma^A \rightarrow  -i \hat\Gamma^A \cr
&& g_{AB} \rightarrow -g_{AB}\cr
&& \psi \rightarrow  i \psi  
\cr
&& 
 \epsilon \rightarrow   \epsilon 
\cr
&& C \rightarrow -i C  
\eea
together with  $\bar\psi  \rightarrow  \bar\psi$ and  $\bar\epsilon \rightarrow - i \bar\epsilon$ which follow from 
the definitions of the Dirac conjugation and the Gamma matrix transformation rule.  
Thus the $SO(4,1)$ twisted and the time reduced theories are equivalent to those from the
$(2,9)$ theory.

\subsection{$SO(5)$ R symmetry group}
The supersymmetry variations and the Lagrangian for the usual Lorentzian M5 brane with $SO(5)$ R symmetry are in our conventions given by
\bea
\delta B_{MN} &=& i \bar\epsilon \Gamma_{MN} \psi\cr
\delta \phi^A &=& i \bar\epsilon \hat\Gamma^A \psi\cr
\delta \psi &=& \frac{1}{12} \Gamma^{MNP} \epsilon H_{MNP} + \Gamma^M \hat\Gamma_A \epsilon \partial_M \phi^A
\eea
and 
\bea
\La &=& -\frac{1}{24} H^{MNP} H_{MNP} - \frac{1}{2} \partial^M \phi^A \partial_M \phi_A + \frac{i}{2} \bar\psi \Gamma^M \partial_M \psi
\eea
Although these variations and the Lagrangian are on the same form as for the case of $SO(4,1)$ R symmetry above, there is no simple relation between the $SO(1,4)$ or $SO(4,1)$ theories and the usual $SO(5)$ theory since there is no natural map from the Dirac conjugate $\bar\psi = \psi^{\dag} \Gamma^0$ to  the Dirac conjugates of  the $SO(1,4)$ or $SO(4,1)$ theories.

\section{Unitarity}
As we have changed signatures, it is important to check unitarity of the theory. To illustrate unitarity, we follow the arguments in \cite{Witten:1989ip}. Let us consider some Lagrangian 
\bea
L &=& \frac{1}{2} g_{ij} \dot q^i \dot q^j + i h_{ij} \psi^{*i} \dot \psi^j
\eea
where $g_{ij}$ and $h_{ij}$ are invertible matrices with inverses $g^{ij}$ and $h^{ij}$. This system can be quantized by imposing the commutation relations
\ben
[q^i,p_j] &=& i \hbar \delta^i_j\label{ett}\\
\{\psi^i,\psi^{j\dag}\} &=& \hbar h^{ij}\label{tva}
\een
where $p_i$ are the conjugate momenta of $q^i$. These have the unitary representations $p_i = - i \hbar \partial/\partial q^i$ irrespectively of the signature of $g_{ij}$, in the sense that the translation operators $U = $exp$ i L^i p_i$ are unitary for any real distances $L^i$, provided the bosonic part of the Lagrangian is real. However, if $g_{ij}$ is indefinite the energy is unbounded from below. For the fermions the situation is the opposite; we see that $h^{ij}$ has to be positive definite to have a unitarity representation of (\ref{tva}). On the other hand we do not encounter negative energy states by filling up the Dirac sea. 

Let us now consider our theory. The bosonic part of our Lagrangian is real although we have indefinite $g_{ij}$. Hence the bosonic part describes a unitary theory. The fermionic part does not however. Here we have
\bea
\{\psi,\psi^{\dag}\} &\sim & \hbar \hat\Gamma^0
\eea
which is indefinite. Hence our 6d theory is non-unitary. This happens for both $SO(1,4)$ and $SO(4,1)$ R symmetry. On the other hand, if the R symmetry is $SO(5)$ the 6d theory is unitary since then we have 
\bea
\{\psi,\psi^{\dag}\} &\sim & \hbar {\mb{I}}
\eea
where ${\mb{I}}$ denotes the $16 \times 16$ identity matrix.

The microscopic structure of an 11d theory is of course unclear, but it seems reasonable to think that such a theory would have two time-directions and global symmetry group $SO(2,9)$, that is broken by an embedding of M5 brane down to $SO(1,5) \times SO(1,4)$. But if we have two time-directions, then time-evolution will be rather different from what we are used to and a new concept should replace that of unitarity, which is based on time evolution with just one time direction. 

Since we are not aware of any formalism with two time directions, let us stick to one time direction. Here we can also find that a unitary theory may appear to be non-unitary if we have one time direction and one space direction, if we interpret the space direction as 'time'. To illustrate this, let us consider an action of a 2-component spinor with $\sigma^3$ the third Pauli matrix,
\bea
S &=& \int d^2 x \psi^{\dag} (i\partial_0 + \sigma^3 \partial_1) \psi
\eea
If we let $x^0$ play the role of time, we quantize the theory by imposing 
\bea
\{\psi^{\dag},\psi\} &=& \hbar
\eea
and we have a unitary representation. But we can also quantize this theory by declaring that $x^1$ is the direction of time evolution, in which case we shall impose the commutation relation
\bea
\{\psi^{\dag},\psi\} &=& \sigma^3
\eea
which has no unitary representation as the matrix $\sigma^3$ is indefinite. One might now speculate that our non-unitary M5 brane theory might appear to be non-unitary for a similar reason that is related to the fact that one time direction of the 11d theory is outside the worldvolume of the M5 brane. 

More concrete statements can be made related to unitarity if we reduce our M5 brane theory along the world-volume time direction. This dimensional reduction gives rise to 5d SYM theory with global symmetry $SO(5) \times SO(1,4)$ and can be exactly mapped to the 5d SYM theory that one would also obtain by reducing 10d SYM theory with $SO(1,9)$ Lorentz symmetry, along time and four space directions. We present the map in full detail in Appendix $B$. As the 10d SYM theory is a unitary theory and the dimensional reduction is a physically consistent procedure, we conclude that there is no problem with our M5 brane theory with $SO(1,4)$ R symmetry after this theory has been reduced along the time direction down to 5d SYM theory. 

Let us finally comment on the issue of convergence of the path integral. If the R symmetry group is $SO(1,4)$, then we have the wrong sign of the kinetic term in the Lagrangian for one of the scalar fields, say $\phi^0$. We may Wick rotate this into $i \phi^0$ to get the right sign kinetic term. We can indeed Wick rotate the R symmetry $SO(1,4)$ including the fermionic part, into the $SO(5)$ R symmetry and get the usual M5 brane theory. But we can also carry on with our $SO(1,4)$ R symmetry, and perform some partial twist of say an $SO(1,p-1)$ subgroup of the R symmetry where $p = 2,3,..$. In this case, the R symmetry will be reduced by the twist with the Lorentz group to $SO(5-p)$. Nevertheless, we can Wick rotate $\phi^0$ into $i\phi^0$ and get the right sign kinetic term. If we do that after the twist, then we get a different theory that can not be related to the familiar M5 brane theory with $SO(5)$ R symmetry.

\section{Superconformal symmetry}
The Lagrangian has not only the usual Poincare supersymmetry, but also a special conformal supersymmetry. We can relax the condition that the supersymmetry parameter is constant, to the condition that it satisfies the superconformal Killing spinor equation \cite{Linander:2011jy}
\bea
D_M \epsilon &=& \frac{1}{6} \Gamma_M \Gamma^N D_N \epsilon
\eea
Once we have done that, we can also admit more general curved six-manifolds where this equation has some solution. The Ricci curvature scalar may be defined by the equation
\bea
\Gamma^{MN} D_M D_N \epsilon &=& -\frac{1}{4}R \, \epsilon
\eea
and $D_M = \partial_M + \omega_M$ is the covariant derivative where $\omega_M$ is the spin connection. 

The Lagrangian is now given by
\bea
\La &=& \La_0 + \La_1
\eea
where
\bea
\La_0 &=& -\frac{1}{24} H_{MNP}^2 + \frac{1}{2} (D_M \phi_A)^2 - \frac{1}{2} \bar\psi \Gamma^M D_M \psi\cr
\La_1 &=& \frac{R}{10} \phi^A \phi_A
\eea
The superconformal symmetry variations can be expressed as 
\bea
\delta &=& \delta_0 + \delta_1
\eea
where
\bea
\delta_0 \phi^A &=& \bar\epsilon \hat\Gamma^A \psi\cr
\delta_0 B_{MN} &=& i \bar\epsilon \Gamma_{MN} \psi\cr
\delta_0 \psi &=& -\frac{i}{12}\Gamma^{MNP} \epsilon H_{MNP} + \Gamma^M \hat\Gamma_A \epsilon D_M \phi^A
\eea
and
\bea
\delta_1 \phi^A &=& 0\cr
\delta_1 B_{MN} &=& 0\cr
\delta_1 \psi &=& \frac{2}{3} \Gamma^M \hat\Gamma_A (D_M\epsilon) \phi^A
\eea
When we vary the Lagrangian, we find it most convenient to bring the variation into the following form,
\bea
\delta \La &=& D_M b^M + \frac{1}{4} D_M H^{MNP} \delta B_{NP} - D_M D^M \phi^A \delta \phi_A 
+\frac{R}{5} \phi^A \delta \phi_A - \bar\psi \Gamma^M D_M \delta \psi
\eea
where the boundary term
\bea
b^M &=& - \frac{1}{4} H^{MNP} \delta B_{NP} + D^M \phi^A \delta\phi_A + \frac{1}{2} \bar\psi \Gamma^M \delta \psi\cr
&=& -\frac{i}{24} \bar\epsilon \Gamma^M \Gamma^{PQR}  \psi H_{PQR} +
\frac{1}{2} \bar\epsilon \hat\Gamma_A \Gamma^M \Gamma^P \psi D_P \phi^A -2 (D^M \bar\epsilon) \hat\Gamma_A \psi \phi^A
\eea
is non-vanishing if the M5 brane has a boundary. If there is no boundary, then the variation is vanishing if the supersymmetry parameter $\epsilon$ is a superconformal Killing spinor. We then find the following superconformal variations,
\bea
\delta_0 \La_0 &=& 4 \bar\psi \hat\Gamma_A (D^N \epsilon) D_N \phi^A\cr
\delta_1 \La_0 &=& - 4 \bar\psi \hat\Gamma_A (D^N \epsilon) D_N \phi^A - \frac{R}{5} \phi^A \bar\epsilon \hat\Gamma_A \psi\cr
\delta_0 \La_1 &=& \frac{R}{5} \phi^A \bar\epsilon \hat\Gamma_A \psi\cr
\delta_1 \La_1 &=& 0
\eea
where we have used the conformal Killing spinor equation and ignore the total derivative contribution $D_M b^M$. Hence $\delta \La = \delta_0 \La_0 + \delta_1 \La_0 + \delta_0 \La_1 + \delta_1 \La_1 = 0$ up to the total derivatives. If we then replace $\epsilon$ by $f \epsilon$ where $f$ is a function on spacetime, then we pick up a variation that is proportional to $\partial_M f$, which is again up to total derivatives. From this we can read off the supercurrent. We only need to consider the last term since this is the only term that can produce something $\sim \partial_M f$. We find that 
\bea
\delta \La &=& j^M \partial_M f
\eea 
where
\bea
j^M &=&- \frac{i}{12} \bar\epsilon \Gamma^{PQR} \Gamma^M \psi H_{PQR} + \bar\epsilon \hat\Gamma_A \Gamma^P \Gamma^M \psi D_P \phi^A + 4 (D^M\bar\epsilon) \hat\Gamma_A \psi \phi^A
\eea
For this computation, we may use the variation
\bea
\delta_1 \psi &=& \frac{2}{3} \Gamma^M \hat\Gamma_A f (D_M\epsilon) \phi^A
\eea
When the equations of motion are satisfied, we will have that the action is stationary under any variation. Hence
\bea
0 = \int d^6 x \sqrt{g} \delta \La = \int j^M D_M f = -\int D_M j^M f 
\eea
and since $f$ is arbitrary, it follows that $D_M j^M = 0$.

\subsection{Coupling to background R symmetry gauge potential}
We introduce a background gauge potential $A_M{}^A{}_B$ and corresponding covariant derivatives 
\bea
D_M \phi^A &=& \nabla_M \phi^A + A_M{}^A{}_B \phi^B\cr
D_M \psi &=& \nabla_M \psi + \frac{1}{4} A_{MAB}\hat\Gamma^{AB} \psi 
\eea
Here $\nabla_M$ is the covariant derivative  of 
 the background geometry. 

We can now find a superconformal Lagrangian by imposing the following Weyl projection
\bea
\frac{1}{2}\Gamma^{MN} \hat\Gamma_A \epsilon F_{MN}{}^A{}_B &=& \hat\Gamma_A \, \epsilon \, P^A{}_B\cr
P_{AB} &=& P_{BA}
\eea
From this, it follows that 
\bea
\frac{1}{2}\Gamma^{MN} \hat\Gamma^{AB}\,  \epsilon F_{MNAB} &=& - \epsilon \, P^A{}_A\cr
\Gamma^{MN} D_M D_N \epsilon &=& -\frac{1}{4}(R+P)\,  \epsilon
\eea
Here we define
\bea
P &=& P^A{}_A
\eea

After we gauge the R symmetry, we find new terms in the variation of the Lagrangian
\bea
\delta_0 \La_0 &=& \cdots - \frac{1}{2} \bar\psi \Gamma^{MN} \hat\Gamma^A \epsilon F_{MNAB} \phi^B = \cdots - \bar\psi \hat\Gamma^A \epsilon P_{AB}\phi^B\cr
\delta_1 \La_0 &=& \cdots -\frac{4}{5} \bar\psi \hat\Gamma_A (\Gamma^{MN} D_M D_N\epsilon) \phi^A = \cdots + \frac{P}{5} \bar\psi\hat\Gamma_A \epsilon \phi^A
\eea
where $\cdots$ are terms of the same form as we had before. We cancel these terms by adding the following terms
\bea
\Delta \La &=& \frac{1}{2}\(\frac{1}{5} \eta_{AB} P - P_{AB}\) \phi^A \phi^B 
\eea
to the Lagrangian.

\subsection{Dimensional reduction along time to 5d SYM}
We assume six-manifold of the form $\mb{R} \times M_5$ with time along $\mb{R}$, and with a rather generic R symmetry gauge field. The natural split of the 6d conformal Killing spinor equation for this analysis will be to write $6 = 1 + 5$, which means that we will assume the following equations
\bea
\Gamma^0 D_0 \epsilon &=& \frac{1}{5} \Gamma^m D_m \epsilon\cr
D_m \epsilon &=& \frac{1}{5} \Gamma_m \Gamma^n D_n \epsilon
\eea
where we also put 
\bea
\partial_0 \epsilon &=& 0
\eea
in order to preserve supersymmetry under the dimensional reduction. 

By dimensional reduction along time, we get the following Lagrangian
\bea
\La_0 &=& \frac{1}{4} F_{mn}^2 + \frac{1}{2} (D_m \phi_A)^2 - \frac{1}{4}[\phi_A,\phi_B]^2\cr
&& - \frac{1}{2} \bar\psi \Gamma^m D_m \psi - \frac{1}{2} \bar\psi \Gamma^0 \hat\Gamma^A [\phi_A,\psi]\cr
\La_1 &=& -\frac{1}{2} (D_0 \phi_A)^2 - \frac{1}{2} \bar\psi \Gamma^0 D_0 \psi + \frac{1}{2} M_{AB} \phi^A \phi^B\cr
\La_2 &=& \frac{i}{6} \epsilon^{ABCDE} A_{0AB} \phi_C [\phi_D,\phi_E]
\eea
where the mass matrix is given by
\bea
M_{AB} &=& \frac{1}{5} \eta_{AB} \(R + P\) - P_{AB}
\eea
The action is invariant under
\bea
\delta \phi_A &=& \bar\epsilon \hat\Gamma_A \psi\cr
\delta A_m &=& i \bar\epsilon \Gamma_m \Gamma_0 \psi\cr
\delta \psi &=& -\frac{i}{2} \Gamma^{mn} \Gamma^0 \epsilon F_{mn} + \Gamma^m \hat\Gamma_A \epsilon D_m \phi^A - \frac{1}{2} \hat\Gamma^{AB} \Gamma^0 [\phi_A,\phi_B]\cr
&& + \Gamma^0 \hat\Gamma_A \epsilon D_0 \phi^A + 4 \Gamma^0 \hat\Gamma^A D_0 \epsilon \phi^A
\eea

To check supersymmetry, we only need to check this for the nonabelian type of terms that involve the curvature corrections. Collecting all such terms, we find the following contributions 
\bea
\delta \La_0 &=& - \frac{3}{2} \bar\psi\hat\Gamma^{AB}D_0 \epsilon [\phi_A,\phi_B] - \bar\psi \hat\Gamma^{AB} \epsilon [\phi_A,D_0\phi_B] - \bar\psi\epsilon [\phi^A,D_0 \phi_A]\cr
\delta \La_1 &=& - \frac{1}{2} \bar\psi\hat\Gamma^{AB}D_0\epsilon [\phi_A,\phi_B] - \bar\psi\hat\Gamma^{AB}\epsilon[\phi_A, D_0\phi_B]
\eea
Then we note
\bea
\hat\Gamma^{AB}\hat\Gamma^{CD} &=& -2 \eta^{AB,CD} + 4 \eta^{BC} \hat\Gamma^{AD} + \hat\Gamma^{ABCD}\cr
D_0 \epsilon &=& \frac{1}{4} \hat\Gamma^{AB}\epsilon A_{0AB}
\eea
and we get
\bea
\delta(\La_0+\La_1) &=& -\frac{1}{2} \bar\psi\hat\Gamma^{ABCD}\epsilon A_{0CD}[\phi_A,\phi_B]
= -\delta \La_2 
\eea

\section{Summary}
The M5 brane Lagrangian is given by 
\bea
\La &=& \La_0 + \La_1
\eea
where 
\bea
\La_0 &=& -\frac{1}{24} H^{MNP} H_{MNP} + \frac{1}{2} \nabla^M \phi^A \nabla_M \phi_A - \frac{1}{2} \bar\psi \Gamma^M \nabla_M \psi,\cr
\La_1 &=& A^M_{AB} \phi^B \partial_M \phi^A + \frac{1}{2} M_{AB} \phi^A \phi^B - \frac{1}{8} \bar\psi \Gamma^M \hat\Gamma^{AB} \psi A_{MAB}
\eea
where $\nabla_M$ is the covariant derivative of the background geometry and  
\bea
M_{AB} &=& \frac{1}{5} \eta_{AB} (P+R) - P_{AB} + A_{M}{}^C{}_A A^{M}{}_{CB}
\eea
We have the superconformal transformations 
\bea
\delta \phi_A &=& \bar\epsilon\hat\Gamma_A\psi\cr
\delta B_{MN} &=& i\bar\epsilon\Gamma_{MN}\psi\cr
\delta\psi &=& -\frac{i}{12}\Gamma^{MNP}\epsilon H_{MNP} + \Gamma^M\hat\Gamma^A\epsilon \partial_M \phi_A + \frac{1}{p}\Gamma^{\mu} \hat\Gamma_A \hat\Gamma^{BC}\epsilon A_{\mu BC} \phi^A + \Gamma^{\mu} \hat\Gamma^A \epsilon A_{\mu AB} \phi^B
\eea
where $P_{AB}$ is a symmetric tensor that we deduce from the curvature of the R-symmetry connection through the Weyl projection
\bea
\frac{1}{2}\Gamma^{MN} \hat\Gamma_A \epsilon F_{MN}{}^A{}_B &=& \hat\Gamma_A \epsilon P^A{}_B
\eea
It would be interesting to see whether one can give $P_{AB}$ a geometric interpretation, perhaps as the Ricci tensor in normal directions to the M5 brane.

By dimensional reduction along time, we can also find a nonabelian generalization
\bea
\La_0 &=& \tr\Bigg(\frac{1}{4} F^{mn} F_{mn} + \frac{1}{2} \nabla^m \phi^A \nabla_m \phi_A - \frac{1}{4} [\phi^A,\phi^B][\phi_A,\phi_B] \cr
&& - \frac{1}{2} \bar\psi \Gamma^m \nabla_m \psi - \frac{1}{2} \bar\psi \Gamma^0 \hat\Gamma^A \psi [\phi_A,\psi]\Bigg)\cr
\La_1 &=& \tr\Bigg(A^M_{AB} \phi^B \partial_M \phi^A + \frac{1}{2} M_{AB} \phi^A \phi^B - \frac{1}{8} \bar\psi \Gamma^M \hat\Gamma^{AB} \psi A_{MAB}\cr
&& +\frac{i}{2} \epsilon^{ABCDE} A_{0AB} \phi_C[\phi_D,\phi_E]\Bigg)
\eea

\section{Six-manifolds on the form $\mb{R}^{1,p-1} \times S^{6-p}$}
We will now restrict ourselves to six-manifolds on the form $\mb{R}^{1,p-1} \times S^{6-p}$ where $p$ can take any of the values $p=1,2,3,4,5,6$. We will subsequently perform a partial topological twist along $\mb{R}^{1,p-1}$, although for $p=1$ this twist cannot be done since the Lorentz group on $\mb{R}$ 
is rather trivial. For our M5 brane theory on $\mb{R}^{1,5}$ we have deliberately chosen the global symmetry group $SO(1,5) \times SO(1,4)$. If we break this symmetry down to $SO(1,p-1) \times SO(6-p) \times SO(1,p-1) \times SO(5-p)$, we can perform a partial twist and identify the two $SO(1,p-1)$ subgroups and declare that the diagonal subgroup of these, times $SO(6-p)$, is the new twisted Lorentz group. Thus after the twist, we have the global symmetry $SO(1,p-1)' \times SO(6-p) \times SO(5-p)_R$. We then first need how the M5 brane spinor in the representation $(4';4)$ of $SO(1,5) \times SO(1,4)$ transforms under the subgroups for the various values of $p$. Here we denote by a prime as in $4'$ the anti-Weyl representation. The supersymmetry parameter is subject to the anti-Weyl projection $\Gamma \epsilon = - \epsilon$. After the split we find the following representations
\bea
p = 1 &&  (4;4)\cr
p = 2 && \(2_{-\frac{i}{2}} \oplus 2'_{+\frac{i}{2}};2_{\frac{i}{2}} \oplus 2_{-\frac{i}{2}}\)\cr
p = 3 && \(2,2;2_{+\frac{1}{2}}\) \oplus \(2,2;2_{-\frac{1}{2}}\)\cr
p = 4 && \(2_{-\frac{1}{2}} \oplus 2'_{+\frac{1}{2}},2 \oplus 2'\)\cr
p = 5 && (4;4)
\eea
where subscripts denote either $SO(1,1)$ or $SO(2)$ charges respectively. Our convention for these charges are $Q^{MN}=-\frac{i}{2}\Gamma^{MN}$ so that for instance $Q^{01} = \pm \frac{i}{2}$ and $Q^{45} = \pm\frac{1}{2}$. After the identification of the $SO(1,p-1)$ groups, these representations become
\bea
p = 1 && (4;4)\cr
p = 2 && (2,2)_{0} \oplus (2',2)_{0} \oplus (2,2)_{-i} \oplus (2',2)_{+i}\cr
p = 3 && (1,2)_{+\frac{1}{2}} \oplus (3,2)_{+\frac{1}{2}} \oplus (1,2)_{-\frac{1}{2}} \oplus (3,2)_{\frac{1}{2}}\cr
p = 4 && 1_{-\frac{1}{2}} \oplus 3^+_{-\frac{1}{2}} \oplus 4_{-\frac{1}{2}} \oplus 4_{+\frac{1}{2}} \oplus 3^-_{+\frac{1}{2}} \oplus 1_{+\frac{1}{2}}\cr
p = 5 && 1 \oplus 5 \oplus 10
\eea
Here $3^+$ refers to a selfdual two-form of $SO(1,3)$. Let us turn to the Weyl projections for the singlet supercharges. First we have the 6d Weyl projection 
\bea
\Gamma^{01}\Gamma^{23}\Gamma^{45} \epsilon &=& -\epsilon
\eea
For $p=2$ we have the singlet representations $(2,2)_{0} \oplus (2',2)_{0}$ i.e. neutral under $SO(1,1)$. For these representations we have
\bea
\Gamma^{01} \hat\Gamma_{0'1'}\epsilon &=& \epsilon
\eea
For $p=3$ we have the singlet representations $ (1,2)_{+\frac{1}{2}} \oplus (1,2)_{-\frac{1}{2}}$ i.e. singlets under $SO(1,2)$. For these representations we have
\bea
\Gamma^{01} \hat\Gamma_{0'1'}\epsilon &=& \epsilon\cr
\Gamma^{12} \hat\Gamma_{1'2} \epsilon &=& \epsilon
\eea
These two projections project onto the singlet state in the tensor product representation of two spin-1/2 representations of $SO(1,2)$. With the gamma matrix representation as below, these two projections amount to 
\bea
(\sigma^3)^{s_0}{}_{s_0'} (\sigma^3)^{t_0}{}_{t'_0} \eta^{s_0't_0'} &=& -\eta^{s_0 t_0}\cr
(\sigma^2)^{s_0}{}_{s_0'} (\sigma^2)^{t_0}{}_{t_0'} \eta^{s_0' t_0'} &=& -\eta^{s_0 t_0}
\eea
The first projection picks states with spins $s_0 + t_0 = 0$, that is either $\|+-\>$ or $\|-+\>$. Then the second projection projects out the even linear combination $\|+-\> + \|-+\>$ leaving us with the singlet state $\|+-\> - \|-+\>$ of $SO(1,2)$. In other words, $\eta^{s_0 t_0} = \epsilon^{s_0 t_0} \eta$ where $\epsilon^{s_0 t_0}$ is the antisymmetric tensor with $\epsilon^{+-} = 1$. This is why we chose the notation $\eta$ for the supersymmetry parameter, in order to not confuse it with the antisymmetric tensor.

After having performed the partial topological twist, we may put the theory on $M_{1,p-1} \times S^{6-p}$ where $M_{1,p-1}$ can be any Lorentzian $p$-dimensional manifold, while preserving a certain amount of supersymmetry. For $p=2$ this will then have applications to the AGT correspondence relating SYM theory on $S^4$ to Toda theory on $M_{1,1}$. For $p=3$ we should expect to find the 3d-3d correspondence with a complex Chern-Simons theory living on $M_{1,2}$. For $p=5$ we have a trivial circle reduction from 6d down to 5d SYM and $p=6$ is flat M5 brane on $\mb{R}^{1,5}$. The case $p=1$ has been considered in  \cite{Kim:2012ava} and in many subsequent papers.

Let us now begin the detailed computations. We split the 6d vector index $M = (\mu,i)$ where $\mu$ lives on $\mb{R}^{1,p-1}$ (and more generally on $M_{1,p-1}$ after the twist) and $i$ lives on $S^{6-p}$. We assume that the background gauge field has no components along $S^{6-p}$,
\bea
A_i &=& 0
\eea
and we require the 6d conformal Killing spinor equation holds along with the conditions that the supersymmetry parameter is constant on $\mb{R}^{1,p-1}$,
\bea
\partial_{\mu} \epsilon &=& 0
\eea
This implies that
\bea
\Gamma^{\mu} D_{\mu} \epsilon &=& \frac{p}{6-p} \Gamma^i D_i \epsilon\cr
D_i \epsilon &=& \frac{1}{6-p} \Gamma_i \Gamma^j D_j \epsilon\cr
D_{\mu} \epsilon &=& \frac{1}{p} \Gamma_{\mu} \Gamma^{\nu} D_{\nu} \epsilon
\eea
and, for $p=2,3,4$,
\bea
P &=& - \frac{p(p-1)}{(6-p)(5-p)} R
\eea
where we have
\ben
\Gamma^{\mu\nu} D_{\mu} D_{\nu} \epsilon = \frac{1}{8} \Gamma^{\mu\nu}\hat\Gamma^{AB} \epsilon F_{\mu\nu AB}
 = -\frac{1}{4}P\, \epsilon\cr
\Gamma^{ij} D_i D_j \epsilon = -\frac{1}{4}R\, \epsilon\label{P}
\een
Let us comment that once we put $\partial_{\mu} \epsilon = 0$ we descend to an ordinary Killing spinor equation on $M_{6-p}$
\bea
D_i \epsilon &=& \frac{1}{4p}  \Gamma_i \Gamma^{\mu} \hat\Gamma^{AB}\epsilon A_{\mu AB}
\eea
For $p=1$ we may instead use the relation
\bea
D_{0} D^{0} \epsilon &=& +\frac{1}{80}R\, \epsilon
\eea
to determine $A_{0,AB}$

We have the curvature condition
\bea
\frac{1}{2} \Gamma^{\mu\nu} \hat\Gamma^{AB} \epsilon F_{\mu\nu AB} &=& -{P}\epsilon  
\eea
Assuming that $p=2,3,4$ we can solve this equation as
\bea
F_{\mu\nu}^{\mu'\nu'} &=& -\frac{2P}{p(p-1)} \delta_{\mu\nu}^{\mu'\nu'}\cr
F_{\mu\nu}^{ab} &=& 0\cr
F_{\mu\nu}^{\mu' a} &=& 0
\eea
if we imposing the Weyl projection
\ben
\frac{1}{p(p-1)} \Gamma^{\mu\nu} \hat\Gamma_{\mu'\nu'} \epsilon &=& \epsilon\label{third}
\een
We find that if we make the assumptions we make, then the curvature $R$ must be constant, and it leads us to consider manifolds on the form $\mb{R}^{1,p-1} \times S^{6-p}$. If $r$ denotes the radius of $S^{6-p}$, then we have
\bea
R &=& \frac{(6-p)(5-p)}{r^2}\cr
P &=& -\frac{p(p-1)}{r^2}
\eea
We further find that 
\bea
P^{\mu'}_{\phantom{a}\nu'}&=& -\frac{p-1}{r^2}\delta^{\mu'}_{\nu'}
\eea
We now proceed to solve the conformal Killing spinor equation on $\mb{R}^{1,p-1}$ with respect to the background gauge field. To this end, it is convenient to introduce the notations
\bea
X_{\mu} &=& \frac{1}{4} \hat\Gamma^{AB}A_{\mu AB}\cr
Y_{\mu} &=& X_{\mu}\epsilon
\eea
The equation we have to solve then reads
\bea
Y_{\mu} &=& \frac{1}{p} \Gamma_{\mu}\Gamma^{\nu} Y_{\nu}
\eea
For $p\neq 1$, we can rewrite this in the form
\ben
Y_{\mu} &=& \frac{1}{p-1} \Gamma_{\mu}{}^{\nu} Y_{\nu}\label{c}
\een
We solve this iteratively in $p$. If we know the solution for $p$, then we can construct the solution for $p+1$. For $p+1$, we have the equations
\ben
Y_{\mu} &=& \frac{1}{p} \Gamma_{\mu}{}^{\nu} Y_{\nu} + \frac{1}{p} \Gamma_{\mu}{}^p Y_p\label{a}\\
Y_p &=& \frac{1}{p} \Gamma_p{}^{\mu} Y_{\mu}\label{b}
\een
Inserting (\ref{b}) into (\ref{a}), we find the equation (\ref{c}). Let us now take $p=2$ which is the lowest value of $p$ for which the conformal Killing spinor on $\mb{R}^{1,p-1}$ is nontrivial. For $p=2$ we get
\bea
Y_{\mu} &=& \Gamma_{\mu}{}^{\nu} Y_{\nu}
\eea
By induction we then find that the most general solution for general $p$ can be expressed as
\ben
Y_{\mu} &=& \Gamma_{\mu}{}^p Y_p\label{first}
\een
for $\mu = 0,\cdots,p-1$.

We also have to satisfy the condition that comes from the curvature by commuting two covariant derivatives as in equation (\ref{P}) that amounts to the condition 
\ben
\Gamma^{\mu\nu} [X_{\mu},X_{\nu}] \epsilon &=& -\frac{1}{2} P\, \epsilon\label{second}
\een
We will now proceed to solve the equations (\ref{first}) and (\ref{second}) while imposing the Weyl projection in (\ref{third}) for various values on $p$.

\subsection{M5 brane on $\mb{R}^{1,0} \times S^5$}
For $p=1$ we find the solution
\bea
A_{0,ab} &=& \(\frac{1}{2r} - \lambda\) \epsilon_{ab}\cr
A_{0,a'b'} &=& \(\frac{1}{2r} + \lambda\) \epsilon_{a'b'}
\eea
where $a=1',2'$ and $a'=3',4'$. These solutions are valid only if we impose the projection 
\bea
\hat\Gamma^{1'2'3'4'} \epsilon &=& -\epsilon
\eea
unless $\lambda = \pm\frac{1}{2r}$ when this projection is not necessary. The Lagrangian is
\bea
\La &=& \La_0 + \(\frac{1}{2r} - \lambda\) \epsilon_{ab} \phi^a \partial_0 \phi^b + \(\frac{1}{2r} + \lambda\) \epsilon_{a'b'} \phi^{a'} \partial_0 \phi^{b'}\cr
&& + \(\frac{15}{8r^2} - \frac{\lambda^2}{2}\) \(\phi_a \phi^a + \phi_{a'}\phi^{a'}\) + \frac{\lambda}{2r} \(\phi_a \phi^a - \phi_{a'} \phi^{a'}\) + \frac{2}{r^2}\phi_{0'} \phi^{0'}\cr
&& -\frac{1}{4r} \bar\psi^- \Gamma^0 \hat\Gamma^{1'2'} \psi^- + \frac{\lambda}{2} \bar\psi^+ \Gamma^0 \hat\Gamma^{1'2'} \psi^+
\eea
where $\psi^{\pm} = \frac{1}{2} \(1 \pm \hat\Gamma^{1'2'3'4'}\) \psi$.

\subsection{M5 brane on $\mb{R}^{1,1} \times S^4$}
For $p=2$ we find the solution
\bea
A_{\mu,\nu'4'} &=& \frac{1}{r}\epsilon_{\mu\nu'}
\eea
where $\epsilon_{01'} = 1$ and antisymmetric, in the sense that $\epsilon_{10'} = -1$. The Weyl projection is
\bea
\Gamma^{01}\hat\Gamma_{0'1'} \epsilon &=& \epsilon
\eea
The M5 brane Lagrangian is
\bea
\La &=& \La_0 + \frac{2}{r} \epsilon^{\mu\nu'} \phi^{4'} \partial_{\mu} \phi_{\nu'} \cr
&&+ \frac{1}{r^2} \(-\phi_{0'}^2 + \phi_{1'}^2 + \phi_{2'}^2 + \phi_{3'}^2\)\cr
&& - \frac{1}{4r} \bar\psi \Gamma^{\mu} \hat\Gamma^{\nu'4'} \psi \epsilon_{\mu\nu'}
\eea

\subsection{M5 brane on $\mb{R}^{1,2} \times S^3$}
For $p=3$ we find the solution 
\bea
A_{\mu,\nu'\lambda'} &=& \frac{1}{r} \epsilon_{\mu\nu'\lambda'}
\eea
where $\epsilon_{01'2'} = 1$ and totally antisymmetric. We have the Weyl projections
\bea
\Gamma^{01} \hat\Gamma_{0'1'} \epsilon &=& \epsilon\cr
\Gamma^{12} \hat\Gamma_{1'2'} \epsilon &=& \epsilon
\eea
The M5 brane Lagrangian is
\bea
\La &=&\La_0 + \frac{1}{r} \epsilon^{\mu\nu'\lambda'} \phi_{\lambda'} \partial_{\mu} \phi_{\nu'} - \frac{1}{8r^2} \epsilon_{\mu\nu'\lambda'} \bar\psi \Gamma^{\mu} \hat\Gamma^{\nu'\lambda'} \psi
\eea

\subsection{M5 brane on $\mb{R}^{1,3} \times S^2$}
For $p=4$ we find the solution
\bea
A_{\mu,\nu'4'} &=& \frac{i}{r} \eta_{\mu\nu'}
\eea
and Weyl projections
\bea
\Gamma^{01} \hat\Gamma_{0'1'} \epsilon &=& \epsilon\cr
\Gamma^{12} \hat\Gamma_{1'2'} \epsilon &=& \epsilon\cr
\Gamma^{23} \hat\Gamma_{2'3'} \epsilon &=& \epsilon
\eea
The M5 brane Lagrangian is
\bea
\La &=& \La_0 + \frac{2i}{r} \phi^{4'} \partial^{\mu} \phi_{\mu'} - \frac{3}{r^2} \phi^{4'} \phi_{4'} - \frac{i}{4r} \bar\psi \Gamma_{\mu} \hat\Gamma^{\mu'4'} \psi
\eea

Here we could not find a real solution for the background gauge potential. The 5d SYM action can be real for R symmetry group $SO(2,3)$ if the signature is $(2,3)$ (section 9.2 in \cite{Hull:2014cxa}). We find that the bosonic part of the action is real once we Wick rotate $\phi^{4'}$ which suggests R symmetry is Wick rotated from $SO(1,4)$ into 
$SO(2,3)$. If we do that Wick rotation of R symmetry then $\hat
\Gamma^{4'}$ shall also be Wick rotated and the full action becomes real on
$R^3 \times S^2$ if the signature is $(2,3)$ with the $S^2$ part timelike.

\section{Partially twisted theory on $\mb{R}^{1,1}\times \mb{R}^4$}
For our gamma matrix conventions for this twist, we refer to Appendix \ref{C1}. On $\mb{R}^{1,1}$ we have the flat metric 
\bea
ds^2 = -e^0 e^0 + e^1 e^1 = -2 e^+ e^- - 2 e^- e^+
\eea
where $e^0 = dx^0$ and $e^1 = dx^1$ and we define  
\bea
e^{\pm} &=& \frac{1}{2} \(e^0 \pm e^1\)
\eea
and $\pm$ denote flat lightcone indices. We define 
\bea
\phi^{\pm} &=& \frac{1}{2} \(\phi^0 \pm \phi^1\)
\eea
and
\bea
\gamma^{\pm} &=& \frac{1}{2}\(\gamma^0 \pm \gamma^1\)
\eea
whose nonvanishing components are $(\gamma^+)^+{}_- = 1$ and $(\gamma^-)^-{}_+ = -1$ respectively. We then have
\bea
\gamma^{+-} &=& -\frac{1}{2}\gamma
\eea

We have the following anti-hermitian $SO(1,1)$ charge generator
\bea
Q &=& \frac{i}{2}\gamma^{01}\cr
Q &=& 2i (\delta^{01})_{\mu}{}^{\nu}
\eea
in the spinor and vector representations. It acts on the vector infinitesimally as 
\bea
\delta \phi_{\mu} &=& -\frac{i}{2} \epsilon_{\kappa\tau} (Q^{\kappa\tau})_{\mu}{}^{\nu} \phi_{\nu}
\eea
which yields
\bea
\delta \phi_{\pm} = \pm \epsilon_{01} \phi_{\pm} =: -i \epsilon_{01} Q \phi_{\pm}
\eea
which shows that $\phi_{\pm}$ carry $SO(1,1)$ charge $Q= \pm i$.

We define twisted spinor components as
\bea
\psi_0^{(\pm)\alpha t} &=& \psi^{\pm\alpha \mp t}\cr
\chi_{\pm}^{(\pm)\alpha t}  &=& \psi^{\pm\alpha \pm t} 
\eea
Here, on the left hand side, stands the twisted spinor fields, and $\pm,0$ without round brackets refers to the twisted $SO(1,1)$ charge. The $(\pm)$ refers to the $SO(4)$ Weyl projection on the Dirac spinor index $\alpha$. On the right hand side stands the untwisted spinor fields, and the $\pm$ there refers to $SO(1,1)$ and $SO(1,1)_R$ charges respectively. Hence the total charge of $\psi_0^{(\pm)\alpha t}$ is zero, while $\chi_{\pm}^{(\pm)\alpha t}$ carry $SO(1,1)$ charges $\pm i$ respectively, just like $\phi_{\pm}$ do. In the sequel we will use the following shorthand notations,
\bea
\psi^{(\pm) \alpha t} &:= & \psi_0^{(\pm)\alpha t}\cr
\chi_{\pm}^{\alpha t} &:=& \chi_{\pm}^{(\pm)\alpha t}
\eea

We define
\bea
D_{\pm} &=& e_{\pm}^{\mu} D_{\mu}
\eea
We have
\bea
g^{\mu\nu} D_{\mu} D_{\nu} &=& -\frac{1}{2} \{D_+,D_-\}
\eea
Using the zweibein to convert $\mu$ into flat space indices $\pm$, we find the following twisted Lagrangian
\bea
\La_{tensor} &=& \frac{1}{16} H_{+-}{}^i H_{+-i} + \frac{1}{16} H_-{}^{ij} H_{+ij} + \frac{1}{16} H_+{}^{ij} H_{-ij} - \frac{1}{24} H^{ijk} H_{ijk}
\eea
\bea
\La_{scalars} &=& -\frac{1}{2} g^{\mu\nu} D_{\mu} \phi_+ D_{\nu} \phi_- - \frac{1}{2} g^{ij} \partial_i \phi_+ \partial_j \phi_- + \frac{1}{2} g^{\mu\nu} \partial_{\mu} \phi^a \partial_{\nu} \phi^a + \frac{1}{2} g^{ij} \partial_i \phi^a \partial_j \phi^a
\eea
\bea
\La_{fermions} &=& \bar\chi_- D_+ \psi^- + \bar\chi_+ D_- \psi^+ + \bar\chi_-\gamma^i D_i \chi_+ + \bar\psi^-\gamma^i D_i \psi^+
\eea
where we define the new Dirac conjugation by  $\bar\psi =\psi^\dagger $ with the reality condition $\bar\psi_{\alpha t}=(\psi^{\alpha t})^* =\psi^{\alpha' t'}C_{\alpha' \alpha} \epsilon_{t' t}$.
The action is invariant under the supersymmetry variations
\bea
\delta B_{+-} &=& -2i \bar\epsilon^- \psi^- - 2i \bar\epsilon^+ \psi^+\cr
\delta B_{\pm i} &=& \pm 2i \bar{\epsilon}^{\mp} \gamma_i \chi_{\pm}\cr
\delta B_{ij} &=& i\bar{\epsilon}^- \gamma_{ij} \psi^- - i\bar{\epsilon}^+\gamma_{ij}\psi^+
\eea
\bea
\delta \phi_+ &=& - 2 \bar\epsilon^+ \chi_+\cr
\delta \phi_- &=& - 2 \bar\epsilon^- \chi_-\cr
\delta \phi^a &=& \bar\epsilon^+ \sigma^a \psi^+ + \bar\epsilon^- \sigma^a \psi^-
\eea
\bea
\delta \psi^{\pm} &=& \epsilon^{\pm} D_{\pm} \phi_{\mp} + \gamma^i \sigma^a \epsilon^{\mp} D_i \phi^a +\frac{i}{4} \gamma^i \epsilon^{\mp} H_{+-i} \mp \frac{i}{12} \gamma^{ijk} \epsilon^{\mp} H_{ijk}\cr
\delta \chi_{\pm} &=& - \sigma^a \epsilon^{\pm} D_{\pm} \phi^a - \gamma^i \epsilon^{\mp} D_i \phi_{\pm} \mp \frac{i}{4} \gamma^{ij} \epsilon^{\pm} H_{\pm ij}
\eea

\section{Partially twisted theory on $M_{1,1} \times \mb{R}^4$}
We introduce the Grassmannian two-space vector field by 
\bea
\chi_\mu =e_\mu^+ \chi_++ e_\mu^- \chi_-
\eea
and a scalar 
\bea
\psi= \psi^+ + \psi^-
\eea
where all the Grassmannian fields are realized in the 8d ($\alpha$,$t$) space.
The supersymmetry parameter is a Grassmannian scalar given by
\bea
\epsilon= \epsilon^+ + \epsilon^-
\eea
For notational convenience  let us introduce 6D Weyl projection on $\chi_\mu$ as
\bea
 \chi^{W}_\mu &=&\frac{1}{2}\(\chi_\mu - \gamma_{(4)}\epsilon_{\mu\nu}\chi^\nu \)\cr
\eea
Then  $\chi_\mu$ is subject to the Weyl projection condition  
\bea
\chi_\mu =\chi^{W}_\mu
\eea
which leads to the relation
\bea
\chi_\mu =-\gamma_{(4)}\epsilon_{\mu\nu}\chi^\nu 
\eea

Using this notation, we find the following twisted Lagrangian
\bea
\La_{tensor} &=& -\frac{1}{8} H^{\mu \nu i} H_{\mu \nu i} - \frac{1}{8} H^{\mu ij} H_{\mu ij} - \frac{1}{24} H^{ijk} H_{ijk}
\eea
\bea
\La_{scalars} &=& \frac{1}{4} \phi_{\mu\nu} \phi^{\mu\nu}+\frac{1}{2}(\nabla_\mu \phi^\mu)^2 + \frac{1}{2}g^{ij} \partial_i \phi_\mu \partial_j \phi^\mu
 + \frac{1}{2} g^{\mu\nu} \partial_{\mu} \phi^a \partial_{\nu} \phi^a + \frac{1}{2} g^{ij} \partial_i \phi^a \partial_j \phi^a
\eea
\bea
\La_{fermions} &=& 2\partial_\mu \bar{\psi}  \chi^\mu 
-\bar{\chi}^\mu \gamma^i  \partial_i \chi_\mu +\frac{1}{2}\bar{\psi} \gamma^i  \partial_i \psi
\eea

The action is invariant under the supersymmetry variations
\bea
\delta B_{\mu\nu} &=& -i \epsilon_{\mu\nu} \, \bar\epsilon \psi = i \epsilon_{\mu\nu} \, \bar\psi \epsilon \cr
\delta B_{\mu i} &=& \ 2 i \bar{\epsilon} \gamma_i \gamma_{(4)} \chi_{\mu}\cr
\delta B_{ij} &=&- i\bar{\epsilon} \gamma_{ij} \gamma_{(4)}\psi  
\eea
\bea
\delta \phi_\mu &=& -  2 \bar\epsilon \chi_\mu\cr
\delta \phi^a &=& \bar\epsilon \sigma^a \psi 
\eea
\bea
\delta \psi &=& -\epsilon\, 
\nabla_\mu \phi^\mu -\gamma_{(4)} \epsilon\,  \epsilon^{\mu\nu} \partial_\mu \phi_{\nu}
+ \gamma^i \sigma^a \epsilon \, \partial_i \phi^a -
\frac{i}{4} \gamma^i \epsilon\,  \epsilon^{\mu\nu} H_{\mu\nu i} + \frac{i}{12} \gamma^{ijk} \gamma_{(4)}\epsilon \, H_{ijk}\cr
\delta \chi_{\mu} &=& \frac{1}{2} \(q_\mu -  \gamma_{(4)}\epsilon_{\mu\nu} q^\nu \) \equiv q^{W}_{\mu}
\eea
where
\bea
q_\mu=  - \sigma^a \epsilon \,  \partial_{\mu} \phi^a - \gamma^i \epsilon \,  \partial_i \phi_{\mu} -  \frac{i}{4} \gamma^{ij}
\gamma_{(4)} \epsilon  H_{\mu ij}
\eea

\section{Partially twisted theory on $M_{1,1} \times {S}^4$}
Using the notation of the previous section, we find the following twisted Lagrangian
\bea
\La_{tensor} &=& -\frac{1}{8} H^{\mu \nu i} H_{\mu \nu i} - \frac{1}{8} H^{\mu ij} H_{\mu ij} - \frac{1}{24} H^{ijk} H_{ijk}
\eea
\bea
\La_{scalars} &=& \frac{1}{4} \phi_{\mu\nu} \phi^{\mu\nu}+\frac{1}{2}(\nabla_\mu \phi^\mu)^2 + \frac{1}{2} g^{ij} \partial_i \phi_\mu \partial_j \phi^\mu
 + \frac{1}{2} g^{\mu\nu} \partial_{\mu} \phi^a \partial_{\nu} \phi^a + \frac{1}{2} g^{ij} \partial_i \phi^a \partial_j \phi^a \cr
&-&  \frac{2}{r} \phi^4 \epsilon^{\mu\nu} \partial_\mu \phi_\nu +\frac{1}{r^2}\(
\phi^\mu\phi_\mu  + \phi^{a'} \phi^{a'}
\)
\eea
\bea
\La_{fermions} &=&  2 \partial_\mu \bar{\psi}  \chi^\mu 
-\bar{\chi}^\mu \gamma^i  D_i \chi_\mu +\frac{1}{2}\bar{\psi} \gamma^i  D_i \psi \cr
&-& \frac{1}{2 r} \bar\psi \gamma_{(4)} \sigma^3    \psi 
\eea
Here, in $\La_{scalars}$, we assume indices range as $a=(a',4)$ for $a'=2,3$. 

The action is invariant under the supersymmetry variations
\bea
\delta B_{\mu\nu} &=& -i \epsilon_{\mu\nu} \, \bar\epsilon \psi = i \epsilon_{\mu\nu} \, \bar\psi \epsilon \cr
\delta B_{\mu i} &=& \ 2 i \bar{\epsilon} \gamma_i \gamma_{(4)} \chi_{\mu}\cr
\delta B_{ij} &=&- i\bar{\epsilon} \gamma_{ij} \gamma_{(4)}\psi  
\eea
and
\bea
\delta \phi_\mu &=& -  2\bar\epsilon \chi_\mu\cr
\delta \phi^a &=& \bar\epsilon \sigma^a \psi 
\eea
where $\phi^2, \phi^3$, and $\phi^4$ are respectively matched with $\sigma^1, \sigma^2$ and $\sigma^3$ with a little abuse of notation.
The fermionic variation beomes
\bea
\delta \psi &=& -\epsilon\, 
\nabla_\mu \phi^\mu -\gamma_{(4)} \epsilon\,  \epsilon^{\mu\nu} \partial_\mu \phi_{\nu}
+ \gamma^i \sigma^a \epsilon \, \partial_i \phi^a \cr
&-&
\frac{i}{4} \gamma^i \epsilon\,  \epsilon^{\mu\nu} H_{\mu\nu i} + \frac{i}{12} \gamma^{ijk} \gamma_{(4)}\epsilon \, H_{ijk}
\cr
&+& \frac{2i}{r}(\gamma_{(4)} \sigma_1 \epsilon \phi^3 -\gamma_{(4)} \sigma_2 \epsilon \phi^2 )
\cr
\delta \chi_{\mu} &=& \frac{1}{2} \( q_\mu -\frac{1}{2} \gamma_{(4)}\epsilon_{\mu\nu} q^\nu \)= {q}^W_\mu
\eea
where
\bea
q_\mu=  - \sigma^a \epsilon \,  \partial_{\mu} \phi^a - \gamma^i \epsilon \,  \partial_i \phi_{\mu} -  \frac{i}{4} \gamma^{ij}
\gamma_{(4)} \epsilon  H_{\mu ij} -\frac{1}{r} \gamma_{(4)} \sigma^3 \epsilon \, \phi_\mu
\eea

The Killing spinor equation reads
\bea
D_i \epsilon = \frac{1}{2 r} \gamma_i \gamma_{(4)} \sigma^3 \epsilon
\eea
whose justification follows from the relation
\bea
- (\bar{\psi} \Gamma^i D_i \epsilon)|_{\chi_\pm=0} = - 4 \bar{\psi} M \epsilon |_{\chi_\pm=0} =\bar\psi \gamma^i D_i \epsilon
\eea
where 
\bea
M= \frac{1}{2r} \Gamma^0 \hat{\Gamma}^{14}
\eea

\section{Partially twisted theory on $M_{1,2} \times S^3$}
For our gamma matrix conventions for this twist, we refer to Appendix \ref{C2}. We introduce a Grassmannian vector field $\psi_\mu$ and scalar field $\psi$
where all the Grassmannian fields are realized in the 4d $(s_1,t_1)$ space and $\mu = 0,1,2$. The supersymmetry parameter is a Grassmannian scalar on $M_{1,2}$ which we denote by $\eta$ which is related to the original supersymmetry parameter by
\bea
\epsilon^{s_0 s_1 s_2| t_0 t_1} &=& \epsilon^{s_0 t_0} \eta^{s_1 t_1} 
\eea
In the twisted theory, the reality condition on any Grassmanian fields $\chi$ becomes
\bea
\bar{\chi}_{s_1 t_1}= (\chi^{s_1 t_1})^* = i \chi^{s'_1 t'_1} \epsilon_{s'_1 s_1}\epsilon_{t'_1 t_1}
\eea 
which basically defines the induced charge conjugation matrix for our twisted theory. In addition, we introduce (for more details we refer to Appendix \ref{C2})
\bea
\gamma^i= \gamma^i \otimes 1
\eea
and
\bea
(\sigma^3, \ \kappa^a)= (1 \otimes \sigma^3, \ 1 \otimes \kappa^a)
\eea

With these preliminaries, we find the following  twisted Lagrangian
\bea
\La_{tensor} &=& - \frac{1}{24} H^{\mu\nu\lambda} H_{\mu\nu\lambda}-\frac{1}{8} H^{\mu \nu i} H_{\mu \nu i} - \frac{1}{8} H^{\mu ij} H_{\mu ij} - \frac{1}{24} H^{ijk} H_{ijk}
\eea
\bea
\La_{scalars} &=& \frac{1}{4} \phi_{\mu\nu} \phi^{\mu\nu}+\frac{1}{2}(\nabla_\mu \phi^\mu)^2 + \frac{1}{2} g^{ij} \partial_i \phi_\mu \partial_j \phi^\mu
 + \frac{1}{2} g^{\mu\nu} \partial_{\mu} \phi^a \partial_{\nu} \phi^a + \frac{1}{2} g^{ij} \partial_i \phi^a \partial_j \phi^a \cr
&+&  \frac{1}{r} \epsilon^{\mu\nu\lambda} \phi_\mu \partial_\nu \phi_\lambda
\eea
where $a=3,4$ and 
\bea
\La_{fermions} &=& - 2 \bar{\psi}^\mu   \sigma^3 \nabla_\mu\psi -\epsilon^{\mu\nu\lambda} \bar\psi_\mu \sigma^3 \partial_\nu \psi_\lambda
-i \bar{\psi} \gamma^i \sigma^3  D_i \psi + i \bar{\psi}^\mu \gamma^i \sigma^3  D_i \psi_\mu
\cr
&+& \frac{3}{2 r} \bar\psi  \sigma^3    \psi + \frac{1}{2 r} \bar\psi^\mu  \sigma^3    \psi_\mu 
\eea

This Lagrangian is invariant under the supersymmetry transformation
\bea
\delta B_{\mu\nu} &=& -2 \epsilon_{\mu\nu\lambda} \, \bar\eta \, \sigma^3 \psi^\lambda 
\cr
\delta B_{\mu i} &=& \ 2 i \bar{\eta} \, \gamma_i \sigma^3 \psi_{\mu}\cr
\delta B_{ij} &=&- 2\bar{\eta} \, \gamma_{ij} \sigma^3 \psi  
\eea
\bea
\delta \phi_\mu &=& -  2i \bar\eta \, \psi_\mu\cr
\delta \phi^a &=& 2i \bar\eta\,  \sigma^3 \kappa^a \psi 
\eea
and 
\bea
\delta \psi &=& -\frac{1}{12}\eta\, \(\epsilon^{\mu\nu\lambda} H_{\mu\nu\lambda}
-\epsilon^{ijk} H_{ijk}
\)
- i\sigma^3 \eta \nabla_\mu \phi^\mu 
-\gamma^i \kappa^a \eta \, \nabla_i \phi^a \cr
&+& \frac{2i}{r} \kappa^a \eta\, \phi^a  \cr
\delta \psi_{\mu} &=& -  \frac{1}{4} \gamma^{ij}
\eta  H_{\mu ij} -  \frac{i}{4} \gamma_{i}\,  \eta  \epsilon_{\mu\nu\lambda} H^{\nu\lambda i}
 +i \kappa^a \eta \,  \nabla_{\mu} \phi^a  + \gamma^i \sigma^3 \eta \,  \nabla_i \phi_{\mu}  -
 i \sigma^3 \eta \epsilon_{\mu}\,^{\nu\lambda}\partial_\nu \phi_\lambda
\eea
To verify the supersymmetry of the action, we note that the 6d conformal Killing spinor equation reduces to the usual Killing spinor equation on $S^3$,
\bea
D_i \eta = -\frac{i}{2r} \gamma_i \,  \eta
\eea

The main application of this twist is to the 3d-3d correspondence. This will be analyzed elsewhere.

\subsection*{Acknowledgement}
DB was
supported in part by 
was supported in part by NRF Grant 2014R1A1A2053737.

\newpage

\appendix

\section{Classification of R symmetry groups for 6d $(2,0)$ theories}\label{A}
We assume Lorentz group $SO(1,5)$ and R symmetry group $SO(q,5-q)$ and attempt to impose the 11d Majorana condition
\bea
\bar\psi &=& \psi^T C
\eea
where we shall define
\bea
\bar\psi &=& \psi^{\dag} \Gamma^0 \hat\Gamma^{1\cdots q}
\eea
Let us assume that we can impose this Majorana condition. We can then pick the Majorana representation for the gamma matrices where the charge conjugation matrix is given by 
\bea
C &=& \Gamma^0
\eea
Since we also have that 
\bea
(\Gamma^M)^T &=& -C\Gamma^M C^{-1}\cr
({\hat{\Gamma}}^A)^T &=& -C\hat\Gamma^A C^{-1}\cr
{\Gamma^M}^{\dag} &=& \Gamma^0 \Gamma^M \Gamma^0\cr
{\Gamma^a}^{\dag} &=& -\Gamma^a \qquad {\mbox{for $a=1,\cdots,q$}}\cr
{\Gamma^{a'}}^{\dag} &=& \Gamma^{a'} \qquad {\mbox{for $a'=q+1,\cdots,5$}} 
\eea
we see that 
\bea
{\Gamma^M}^* &=& \Gamma^M\cr
{\Gamma^a}^* &=& -\Gamma^a\cr
{\Gamma^{a'}}^* &=& \Gamma^{a'}
\eea
The Majorana condition becomes
\bea
\psi^{\dag} \Gamma^{1\cdots q} &=& (-1)^q \psi^T
\eea
Applying transpose on both sides, we get
\bea
C\Gamma^{q\cdots 1}C^{-1} \psi^* &=& \psi
\eea
Using $C=\Gamma^0$ we get
\bea
(-1)^{q+1} \Gamma^{q\cdots 1} \psi^* &=& \psi
\eea
Applying $\Gamma^{1\cdots q}$ on both sides, we get
\bea
\psi^* &=& -\Gamma^{1\cdots q} \psi
\eea
If we complex conjugate again, we get
\bea
{\psi^*}^* = -(-1)^q \Gamma^{1...q} \psi^* = (-1)^q ({\Gamma^{1\cdots q}})^2 \psi
\eea
Now we use that 
\bea
({\Gamma^{1\cdots q}})^2 &=& (-1)^{\frac{q(q+1)}{2}}
\eea
We then get
\bea
{\psi^*}^* &=& (-1)^{\frac{q(q-1)}{2}} \psi
\eea
This is consistent for 
\bea
q(q-1) &\in & 4\mb{Z}
\eea
Solutions are $q = 0,1,4,5$ and correspond to $SO(5)$, $SO(1,4)$, $SO(4,1)$ and $SO(5,0)$.

\section{A map from 6d to 10d Weyl projections}\label{B}
To find the non-Abelian generalization, we first put $r = \infty$. We wish to relate the theory with the dimensional reduction of SYM on $\mb{R}^{1,9}$, dimensionally reduced down to $R^5$. For this SYM we have the Weyl projections
\bea
-i \Gamma^0 \zeta &=& \zeta\cr
-i \Gamma^0 \omega &=& \omega
\eea
for the spinor field and the supersymmetry parameter respectively. These will be related by a unitary transformation to our original variables as
\bea
\psi &=& U \zeta\cr
\epsilon &=& U^{\dag} \omega
\eea
where
\bea
U &=& \frac{1}{\sqrt{2}} \(1 + i \Gamma^0 \Gamma\)
\eea
which has the properties
\bea
U U^{\dag} &=& 1\cr
U^2 &=&  i \Gamma^0 \Gamma\cr
U \Gamma^0 &=& \Gamma^0 U^{\dag}\cr
U \Gamma^m &=& \Gamma^m U\cr
U \Gamma^A &=& \Gamma^A U^{\dag}
\eea
We define
\bea
\bar\epsilon &=& \epsilon^{\dag} \Gamma^0 \Gamma^{0'}\cr
\bar\omega &=& \omega^{\dag} \hat\Gamma^{0'}
\eea
and so we also have the relations
\bea
\bar\epsilon &=& \bar\omega \Gamma_0 U\cr
\bar\psi &=& \bar\zeta \Gamma_0 U^{\dag}
\eea
In terms of these new spinor variables, we get
\bea
\delta \phi_A &=& i \bar\omega \hat\Gamma_A \zeta\cr
\delta A_m &=& i \bar\omega \Gamma_m \zeta\cr
\delta \zeta &=&  \frac{1}{2} \Gamma^{mn} \omega F_{mn} + \Gamma^m \hat\Gamma^A \omega \partial_m \phi_A
\eea
If we now also flip the sign of the matter fields $\phi_A$, 
we find the standard supersymmetry variations of (1+9)d SYM reduced to 5d, for which we have the non-Abelian generalization that is obtained by substituting ordinary derivative with gauge covariant derivative $D_m = \partial_m - i [A_m,\bullet]$ in the adjoint representation, and by adding one commutator term
\bea
\delta' \zeta &=& -\frac{i}{2} \hat\Gamma^{AB} \omega [\phi_A,\phi_B]
\eea
We can then transform this term back into our original, M5 brane adapted, variables and get
\bea
\delta' \psi &=&  -\frac{1}{2} \hat\Gamma^{AB} \Gamma^0 \epsilon [\phi_A,\phi_B]
\eea

Likewise the non-Abelian Lagrangian is in the new variables given by the standard SYM Lagrangian
\bea
\La &=& \frac{1}{4} F^{mn} F_{mn} + \frac{1}{2} D^m \phi^A D_m \phi_A - \frac{1}{4} [\phi^A,\phi^B][\phi_A,\phi_B]\cr
&& -\frac{i}{2} \bar{\zeta} \Gamma^m D_m \zeta - \frac{1}{2} \bar\zeta \hat\Gamma^A [\phi_A,\zeta]
\eea
that in the M5 brane adapted variables translates into
\bea
\La_0 &=& \frac{1}{4} F^{mn} F_{mn} + \frac{1}{2} D^m \phi^A D_m \phi_A - \frac{1}{4} [\phi^A,\phi^B][\phi_A,\phi_B]\cr
&& - \frac{1}{2} \bar{\psi} \Gamma^m D_m \psi - \frac{1}{2} \bar\psi \Gamma^0 \hat\Gamma^A [\phi_A,\psi] 
\eea

\section{Gamma matrix conventions for partial topological twists}\label{C}
When we perform the partial topological twisting we find it convenient to choose gamma matrices according to the dimension of the manifold over which we obtain the scalar supercharges after the twist. 

\subsection{Gamma matrices for the 2d-4d split}\label{C1}
We choose the $SO(1,1)$ gamma matrices $\gamma^{\mu}$ as 
\bea
\gamma^0 &=& i\sigma^2\cr
\gamma^1 &=& \sigma^1
\eea
and we define the $SO(1,1)$ chirality matrix as
\bea
\gamma_{(2)} = \gamma^{01} = \sigma^3
\eea
We have
\bea
(\gamma^{\mu})^T &=& -\epsilon \gamma^{\mu} \epsilon^{-1}\cr
\gamma_{(2)}^T &=& -\epsilon \gamma_{(2)} \epsilon^{-1}
\eea
where $\epsilon = i \sigma^2$. 

We then choose the 11d gamma matrices as
\bea
\Gamma^{\mu} &=& \gamma^{\mu} \otimes 1 \otimes 1 \otimes 1\cr
\Gamma^i &=& \gamma_{(2)} \otimes \gamma^i \otimes 1 \otimes 1\cr
\hat\Gamma^{\mu'} &=& \gamma_{(2)} \otimes \gamma_{(4)} \otimes \gamma^{\mu'} \otimes 1\cr
\hat\Gamma^a &=& \gamma_{(2)} \otimes \gamma_{(4)} \otimes \gamma_{(2)} \otimes \sigma^a
\eea
We let indices range as $\mu = \mu' = 0,1$, $i=1,2,3,4$ and $a=1,2,3$. We then find that the 6d chirality matrix becomes
\bea
\Gamma &=& \gamma_{(2)} \otimes \gamma_{(4)} \otimes 1 \otimes 1
\eea
where we define the $SO(4)$ hermitian chirality matrix as
\bea
\gamma_{(4)} = \gamma^{1234}
\eea
The 6d Weyl condition amounts to 
\bea
\(\gamma_{(2)} \otimes \gamma_{(4)}\otimes 1 \otimes 1\) \psi &=& \psi
\eea
The 11d charge conjugation matrix is 
\bea
C_{11d} &=& \epsilon \otimes C \otimes \sigma^1 \otimes \epsilon
\eea
which is such that 
\bea
C_{11d}^T &=& -C_{11d}\cr
(\Gamma^M)^T &=& -C_{11d} \Gamma^M C_{11d}^{-1}\cr
(\hat\Gamma^A)^T &=& -C_{11d} \Gamma^A C_{11d}^{-1}
\eea
We then have $C^T = -C$ and $\epsilon^T = -\epsilon$. An explicit realization of $SO(4)$ gamma matrices is
\bea
\gamma^{1,2,3} &=& \sigma^{1,2,3} \otimes \sigma^2\cr
\gamma^4 &=& 1\otimes \sigma^1
\eea
and 
\bea
C &=& \epsilon \otimes 1
\eea
Then 
\bea
(\gamma^i)^T &=& C \gamma^i C^{-1}
\eea
Also, if we define
\bea
\gamma_{(4)} = \gamma^{1234} = 1\otimes \sigma^3
\eea
then 
\bea
\gamma_{(4)}^T &=& C \gamma_{(4)} C^{-1}
\eea

We will use spinor indices as follows,
\bea
\psi^{s_0 \alpha t_0 t_1}
\eea
Thus if we write out all spinor indices, we have for instance
\bea
C_{11d} &=& \epsilon_{s_0 s_0'} C_{\alpha\beta} \sigma^1_{t_0 t_0'} \epsilon_{tt'}
\eea
We have that 
\bea
C_{\alpha\beta} &=& -C_{\beta\alpha}\cr
\gamma^i_{\alpha\beta} &=& -\gamma^i_{\beta\alpha}\cr
\gamma^{ij}_{\alpha\beta} &=& \gamma^{ij}_{\beta\alpha}\cr
\gamma^{ijk}_{\alpha\beta} &=& \gamma^{ijk}_{\beta\alpha}
\eea
where we define $\gamma^i_{\alpha\beta} := C_{\alpha\gamma} (\gamma^i)^{\gamma}{}_{\beta}$.

We define
\bea
(\gamma_{(2)})^s{}_t &=& \(\begin{array}{cc}
1 & 0\\
0 & -1
\end{array}\)
\eea 
and 
\bea
(\gamma_{(2)})_{st} &=& \(\begin{array}{cc}
0 & -1\\
-1 & 0
\end{array}\)
\eea

We denote the twisted $SO(1,1)$ neutral spinor components as
\bea
\psi^{\alpha t_1}
\eea
In addition to these, we have the twisted $SO(1,1)$ charged spinor components
\bea
\chi^{\alpha t_1}
\eea
which carry the $SO(1,1)$ charge according to their $SO(4)$ chirality. 

In total we have $8$ neutral (denoted as $\psi$) and $8$ charged (denoted as $\chi$) spinor components. The supersymmetry parameters are neutral under $SO(1,1)$. We denote these as 
\bea
\epsilon^{\alpha t_1}
\eea
which has $4 \times 2 = 8$ real components. In other words, we have $8$ real supercharges.

\subsection{Gamma matrices for the 3d-3d split}\label{C2}
We choose 11d gamma matrices as ($\mu = 0,1,2$, $i=3,4,5$, $A=0',1',2',3',4'$)
\bea
\Gamma^{\mu} &=& \gamma^{\mu} \otimes 1 \otimes \sigma^2 \otimes 1\cr
\Gamma^i &=& 1 \otimes \gamma^i \otimes \sigma^1 \otimes 1\cr
\hat\Gamma^A &=& 1 \otimes 1 \otimes \sigma^3 \otimes \gamma^A
\eea
where $\gamma^{\mu} = (i\sigma^2,\sigma^1,\sigma^3)$ and $\gamma^i = (\sigma^3,\sigma^1,\sigma^2)$ and where we choose $\gamma^A$ as follows
\bea
\gamma^0 &=& i\sigma^2 \otimes \sigma^3\cr
\gamma^1 &=& \sigma^1 \otimes \sigma^3\cr
\gamma^2 &=& \sigma^3 \otimes \sigma^3\cr
\gamma^3 &=& 1 \otimes \sigma^2\cr
\gamma^4 &=& 1 \otimes \sigma^1
\eea
 and we may use the notation
\bea
\gamma^{\mu'} &=& \gamma^{\mu'} \otimes \sigma^3\cr
\gamma^a &=& 1\,\, \otimes \, \kappa^a
\eea
for $\mu'=0,1,2$ and $a=3,4$. We have
\bea
(\gamma^A)^T &=& C\gamma^A C^{-1}\cr
C^T &=& -C
\eea
where
\bea
C &=& \epsilon \otimes \sigma^1
\eea
The 11d charge conjugation matrix is 
\bea
C_{11d} &=& \epsilon \otimes \epsilon \otimes \sigma^1 \otimes C
\eea
which is antisymmetric 
\bea
C_{11d}^T &=& -C_{11d}
\eea
We expand the spinor as 
\bea
\psi^{s_0 s_1 s_2 t_0 t_1} &=& \epsilon^{s_0 t_0} \psi^{s_1 t_1} + (\gamma^\mu)^{s_0 t_0} \psi_\mu^{s_1 t_1}
\eea
Here $\psi^{s_1 \pm}$ transform in the representation $(1,2)_{\pm}$ and $\psi^{s_1 \pm}_\mu$ in the representation $(3,2)_{\pm}$ of $SO(1,2) \times SO(3) \times SO(2)_R$. Note that $s_2$ is determined by the 6d Weyl projection. We have
\bea
\Gamma^{01} &=& (\sigma^3)^{s_0}{}_{s'_0}\cr
\Gamma^{23} &=& -i (\sigma^3)^{s_0}{}_{s'_0} (\sigma^3)^{s_1}{}_{s'_1} (\sigma^3)^{s_2}{}_{s'_2}\cr
\Gamma^{45} &=& i (\sigma^3)^{s_1}{}_{s'_1}
\eea
Then 
\bea
\Gamma = \Gamma^{01}\Gamma^{23}\Gamma^{45} = (\sigma^3)^{s_2}{}_{s'_2}
\eea
We conclude that $s_2$ gives the 6d chirality of the spinor so that this number is fixed by the spinor. For $\psi$ we have $s_2 = +$ and for the supersymmetry parameter $\eta$ we have $s_2 = -$.

\section{Untwisted Fierz identity}\label{D}
We use 11d gamma matrices that we split them into two groups, $\Gamma^M$ and $\hat\Gamma^A$ where $M=0,1,2,3,4,5$ is for spacetime and 
$A=0',1',2',3',4'$ is for $SO(1,4)$ R symmetry. We thus assume that $\{\Gamma^M,\hat\Gamma^A\} = 0$ as part of the 11d Clifford algebra. We define the 6d chirality matrix 
\bea
\Gamma &=& \Gamma^{012345}
\eea
For two negative chirality spinors $\Gamma \epsilon = -\epsilon$ and $\Gamma \eta = -\eta$, we have the following Fierz identity,
\bea
\epsilon \bar \eta - \eta \bar\epsilon &=& \frac{1}{8} \[-(\bar\eta \Gamma^M \epsilon)\Gamma_M + (\bar\eta\Gamma^M\hat\Gamma^A\epsilon)\Gamma_M \hat\Gamma_A\] \frac{1}{2}\(1+\Gamma\)\cr
&& - \frac{1}{192} (\bar\eta\Gamma^{MNP}\hat\Gamma^{AB}\epsilon)\Gamma_{MNP}\hat\Gamma_{AB}
\eea
We have the following gamma matrix identities,
\bea
\Gamma^{MNP}\Gamma_Q \Gamma_{NP} &=& -20\delta^M_Q - 4 \Gamma^M{}_Q\cr
\Gamma^{PMN} \Gamma_{QRS} \Gamma_{MN} &=& 4 \Gamma^P{}_{QRS} + 12 \delta^P_{[Q} \Gamma_{RS]}\cr
\hat\Gamma_A \hat\Gamma^B \hat\Gamma^A &=& -3 \hat\Gamma^B\cr
\hat\Gamma_A \hat\Gamma^{BC} \hat\Gamma^A &=& \hat\Gamma^{BC}
\eea


\begin{thebibliography}{99}


\bibitem{Bergshoeff:1999db} 
  E.~Bergshoeff, E.~Sezgin and A.~Van Proeyen,
  ``(2,0) tensor multiplets and conformal supergravity in D = 6,''
  Class.\ Quant.\ Grav.\  {\bf 16}, 3193 (1999)
  [hep-th/9904085].

\bibitem{Cordova:2013bea} 
  C.~Cordova and D.~L.~Jafferis,
  ``Five-Dimensional Maximally Supersymmetric Yang-Mills in Supergravity Backgrounds,''
  arXiv:1305.2886 [hep-th].

\bibitem{Festuccia:2011ws} 
  G.~Festuccia and N.~Seiberg,
  ``Rigid Supersymmetric Theories in Curved Superspace,''
  JHEP {\bf 1106}, 114 (2011)
  [arXiv:1105.0689 [hep-th]].



\bibitem{Cordova:2013cea} 
  C.~Cordova and D.~L.~Jafferis,
  ``Complex Chern-Simons from M5-branes on the Squashed Three-Sphere,''
  arXiv:1305.2891 [hep-th]. 

\bibitem{Kim:2014kta} 
  J.~Kim, S.~Kim, K.~Lee and J.~Park,
  ``Super-Yang-Mills theories on $S^{4} \times \mathbb{R}$,''
  JHEP {\bf 1408}, 167 (2014)
  [arXiv:1405.2488 [hep-th]].

\bibitem{Kim:2012ava}
  H.~C.~Kim and S.~Kim,
  ``M5-branes from gauge theories on the 5-sphere,''
  JHEP {\bf 1305} (2013) 144
  [arXiv:1206.6339 [hep-th]].

\bibitem{Alday:2009aq}
  L.~F.~Alday, D.~Gaiotto and Y.~Tachikawa,
  ``Liouville Correlation Functions from Four-dimensional Gauge Theories,''
  Lett.\ Math.\ Phys.\  {\bf 91} (2010) 167
  [arXiv:0906.3219 [hep-th]].
  
\bibitem{Gran:2014lia}
  U.~Gran, H.~Linander and B.~E.~W.~Nilsson,
  ``Off-shell structure of twisted (2,0) theory,''
  JHEP {\bf 1411} (2014) 032
  [arXiv:1406.4499 [hep-th]].

\bibitem{Anderson:2013hpa} 
  L.~Anderson and H.~Linander,
  ``The trouble with twisting (2,0) theory,''
  JHEP {\bf 1403}, 062 (2014)
  [arXiv:1311.3300 [hep-th]].

\bibitem{Linander:2011jy}
  H.~Linander and F.~Ohlsson,
  ``(2,0) theory on circle fibrations,''
  JHEP {\bf 1201} (2012) 159
  [arXiv:1111.6045 [hep-th]].

\bibitem{Hull:1999mt}
  C.~M.~Hull and R.~R.~Khuri,
  ``World volume theories, holography, duality and time,''
  Nucl.\ Phys.\ B {\bf 575} (2000) 231
  [hep-th/9911082].

\bibitem{Hull:2014cxa} 
  C.~M.~Hull and N.~Lambert,
  ``Emergent Time and the M5-Brane,''
  JHEP {\bf 1406}, 016 (2014)
  [arXiv:1403.4532 [hep-th]].

\bibitem{Witten:1989ip} 
  E.~Witten,
  ``Quantization of {Chern-Simons} Gauge Theory With Complex Gauge Group,''
  Commun.\ Math.\ Phys.\  {\bf 137}, 29 (1991).


\bibitem{Pestun:2007rz} 
  V.~Pestun,
  ``Localization of gauge theory on a four-sphere and supersymmetric Wilson loops,''
  Commun.\ Math.\ Phys.\  {\bf 313}, 71 (2012)
  [arXiv:0712.2824 [hep-th]].


\end{thebibliography}
\end{document}